\documentclass[aps,prd,twocolumn,groupedaddress,showpacs,showkeys]{revtex4}
\usepackage{latexsym,epsfig,amsmath,amssymb}
\usepackage{amsmath}
\usepackage{amsfonts}
\usepackage{epsfig}
\usepackage{verbatim}

\usepackage{bm}

\def\shiftdown#1{#1\llap{\lower.04ex\hbox{#1}}}

\newcommand{\beaa}{\begin{eqnarray*}} 
\newcommand{\enaa}{\end{eqnarray*}}
\newcommand{\bea}{\begin{eqnarray}}
\newcommand{\ena}{\end{eqnarray}}
\newcommand{\be}{\begin{eqnarray}} 
\newcommand{\eq}{\begin{eqnarray}} 
\newcommand{\en}{\end{eqnarray}}
\newcommand{\ra}{\rangle}

\begin{document}

\title{Two-photon decay of heavy hadron molecules} 

\author{
Tanja Branz, 
Thomas Gutsche, 
Valery E. Lyubovitskij
\footnote{On leave of absence from the
Department of Physics, Tomsk State University,
634050 Tomsk, Russia}
\vspace*{1.2\baselineskip}}

\affiliation{Institut f\"ur Theoretische Physik,
Universit\"at T\"ubingen,\\
Kepler Center for Astro and Particle Physics, \\
Auf der Morgenstelle 14, D--72076 T\"ubingen, Germany
}

\date{\today}

\begin{abstract} 

We discuss the two-photon decay width of heavy hadron molecules and
study the dependence on the constituent meson masses
and on the binding energy. In addition finite size effects due to the extended structure
of the bound state are shown to have a strong influence on the predictions
for this decay width.
\end{abstract}



\maketitle

Conventional quark-antiquark constituent quark model descriptions of the meson spectrum
are certainly incomplete and sometimes inadequate to match experimental
observation. As a complementary or even alternative approach new structure
assumptions have been invoked in the past. This concerns for example glueballs,
quark-gluon hybrids, tetraquark configurations or even an admixture of some of these states
with an ordinary quark-antiquark meson as suggested by direct applications of
(as in lattice simulations) or in models motivated by
the underlying theory of QCD~\cite{Godfrey:2008nc,Klempt:2007cp}. But also conventional configurations  
can populate the meson spectrum as quasi-nuclear nucleon-antinucleon bound states~\cite{Dover:1990kn}
or hadronic molecules, bound states of two or more mesons~\cite{Guo:2008zg,Faessler:2007gv,Branz:2007xp}.

The idea that mesons can also form mesonic bound states traces back to the original
work of~\cite{Weinstein:1982gc,Barnes:1985cy} with a first systematic study for
the meson spectrum given in~\cite{Tornqvist:1993ng}.
The concept of hadronic molecules is often invoked when observed
meson masses are rather close to a respective two-meson threshold and when dynamical arguments
can be found which suggest binding in these systems.
Experimental candidates for these type of states are especially discussed
in light of the experimental advances in the heavy flavor sector. A prominent
example is the $X(3872)$ where a molecular interpretation is suggested~\cite{Voloshin:2003nt}. 

A full interpretation of an observed meson resonance is not only based on a consistent description
of its mass and $J^{PC} (I^G)$ quantum numbers, but also on a possible explanation
of the observed decay and production modes of this respective
state. Here the electromagnetic interaction plays a leading role since
it is well understood even on a effective hadronic level and can be treated perturbatively.
In particular, two-photon or, in general, radiative decays are considered diagnostic
tools which are sensitive to the inner structure of the short-lived resonance~\cite{Barnes:1985cy}.

The purpose of the present manuscript is the discussion of 
the two-photon decays of heavy hadron molecules.
Our aim is to work out the influence of the binding energy
and the constituent meson masses on quantities like couplings
(of the molecular state to its constituent mesons) and electromagnetic
decay amplitudes. Under certain circumstances, as for a local theory, the results are analytical
and the dependence on the above mentioned quantities can be shown explicitly.
Especially ratios of two-photon decay rates of bound states, also in different flavor sectors,
provide a simple and clear estimate whether a molecular structure is likely or not.

For the physical or nonlocal case, which effectively models finite size effects of
hadronic molecules, results can deviate sizably from the limiting case of
a local theory.
Here we find that, in general, the two-photon decay properties of heavy bound states are more sensitive to finite
size effects than hadronic molecules of light mesons.

Several heavy meson and baryon states are considered good candidates for a molecular structure
among them the $D_{s0}^\ast(2317)$, $D_{s1}(2460)$, $X(3872)$, $Y(3940)$, $Y(4140)$, $Y(4660)$ and $\Lambda_c(2940)$.
For a review of the situation in the heavy meson sector we refer to~\cite{Godfrey:2008nc,Swanson:2005tq,Guo:2008zg}. 

For the present study of two-photon decays of heavy meson molecules we choose the framework of a phenomenological
Lagrangian approach developed in~\cite{Faessler:2007gv,Ivanov:1996pz}. The bound state structure of hadronic
molecules is set up by the compositeness condition~\cite{Weinberg:1962hj,Ivanov:1996pz}
which implies that the renormalization constant $Z$
of the hadron wave function is set equal to zero or that the hadron exists 
as a bound state of its constituents. Decay processes of hadronic bound states are described by the coupling of
the final state particles via meson-loop diagrams to the constituents of the
molecular state~\cite{Branz:2009yt,Faessler:2007gv,Branz:2007xp}. 
First calculations for the two-photon decay widths of heavy hadron molecules as 
$Y(3940)= \{D^\ast D^{\ast \dagger}\}$ 
and $Y(4140) = \{ D^\ast_s D^{\ast \dagger}_s \}$ 
have been performed in~\cite{Branz:2009yt}. 

In this note we concentrate on hadronic bound states in the 
meson sector.
We further restrict to $S$-wave hadron molecules with  
quantum numbers $J^{PC} = 0^{++}$ whose constituents are pseudoscalar or scalar 
charmed/bottom mesons.
The coupling of the scalar molecular state $H$ to its constituents is
expressed by the phenomenological Lagrangian: 
\eq\label{LY1} 
{\cal L}_H(x) = g_H H(x) J_H(x) 
\en 
where $g_H$ is the coupling constant,  
$H$ is the field representing the hadronic molecule and 
$J_H(x) = \phi^\dagger(x) \phi(x)$ is the current carrying the quantum numbers 
of the molecule composed of the constituent mesons identified with the field $\phi$.
A nonlocal extension, which corresponds to the physical case of a hadronic molecule
with a finite size, can easily be achieved by 
insertion of a vertex function $\Phi_H(y^2)$ 
into the current $J_H(x)$: 
\eq 
J_H(x) = \int d^4y \, \Phi_H(y^2) \, \phi^\dagger(x+y/2) \, \phi(x-y/2)
\en  
where $y$ is the relative Jacobi coordinate. 
Here $\Phi_H(y^2)$ is the correlation function describing the 
distribution of the constituents inside the molecular state $H$. 
In the present evaluations we use 
a Gaussian form with $\Phi_H(y^2)$: 
$\tilde\Phi_H(p_E^2/\Lambda_{H_i}^2) \doteq \exp( - p_E^2/\Lambda_{H_i}^2)$, 
where $p_{E}$ is the Euclidean Jacobi momentum and $\Lambda_{H_i}$
is a size parameter.

\begin{figure}
\centering{\
\epsfig{figure=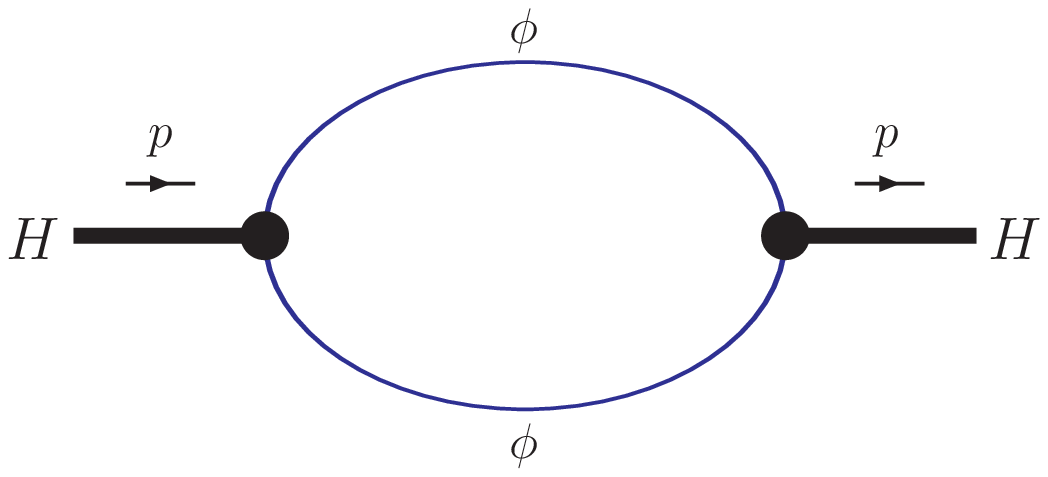,scale=.5}}
\caption{Mass operator of the hadronic molecule $H = (\phi\phi^\dagger)$.}
\label{fig1}

\centering{\
\epsfig{figure=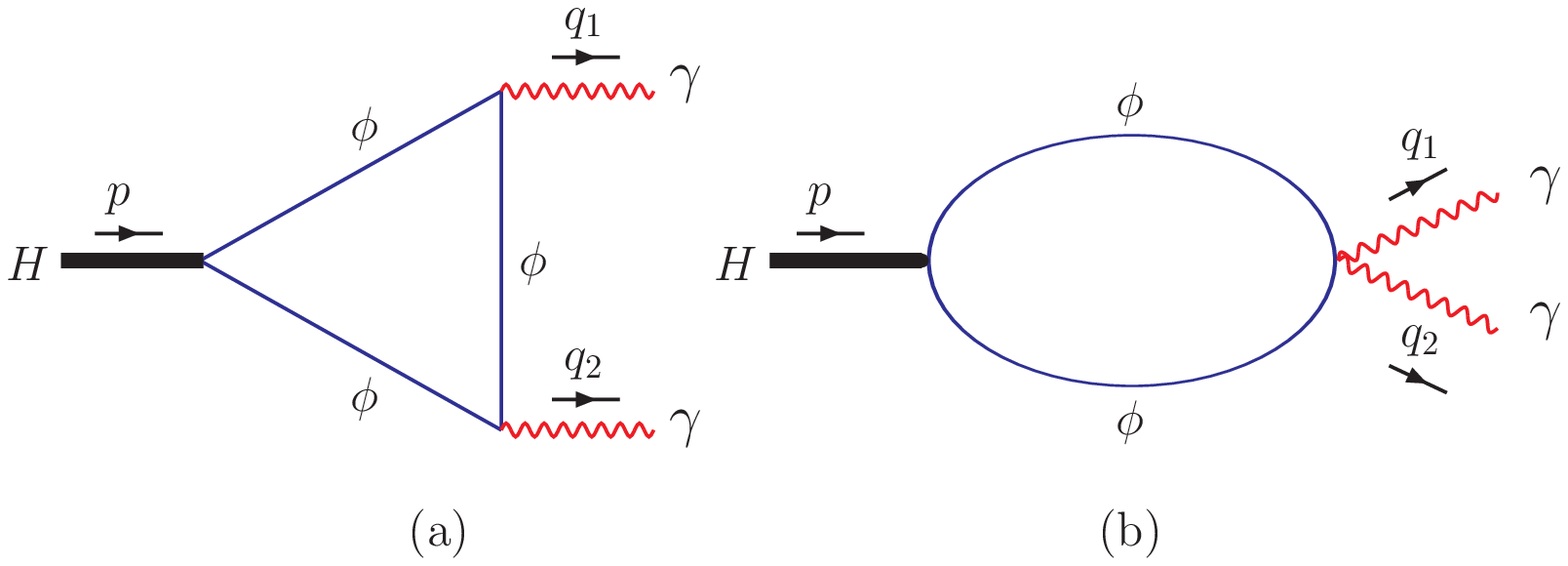,scale=.5}}
\caption{Diagrams contributing to the 
$H \to \gamma\gamma$ decay.} 
\label{fig2}
\end{figure}

The coupling constant $g_H$ is determined by the compositeness 
condition~\cite{Weinberg:1962hj,Ivanov:1996pz}, where $Z = 1 - g_H^2 \Sigma^\prime(M^2) = 0$. Here, $\Sigma^\prime(M^2) = d\Sigma(p^2)/dp^2|_{p^2=M^2}$ 
is the derivative of the mass 
operator $\Sigma(p^2)$ illustrated in Fig.~\ref{fig1}, where $M$ is the mass of the hadronic molecule. 
The generic expression for the mass operator of a bound state of two pseudoscalars is e.g. given in \cite{Branz:2007xp}, 
\eq
\Sigma(p^2)=\int\frac{d^4k}{(2\pi)^4i}\widetilde\Phi(-k^2)S\big(k+\frac p2\big)S\big(k-\frac p2\big)\,,
\en
where $S(k)=(m^2-k^2-i\epsilon)^{-1}$ 
denotes the free propagator of the constituent meson.

In the following we first restrict to the case of a local coupling of the $H$ state to its respective
constituents. In case of a local interaction the vertex function is replaced by
$\tilde\Phi_H(-k^2) = \exp(k^2/\Lambda_{H_i}^2)\equiv 1$. Then the coupling $g_H$
can be expressed in the analytical form \cite{Branz:2007xp,Hanhart:2007wa}: 
\eq 
g^{-2}_H = \frac{1}{(8 \pi m \zeta)^2} \, 
\biggl[ \frac{\beta(\zeta)}{\sqrt{1-\zeta^2}} - 1 \biggr]\,,
\en  
where $\zeta = M/(2m)$, $\beta(\zeta) = {\rm arcsin}(\zeta)/\zeta$ 
and $m$ is the mass of the constituent meson. 

Inclusion of the electromagnetic interaction in a gauge invariant way is discussed in
our previous works~\cite{Branz:2007xp,Faessler:2003yf}. In the local approximation we deal with
the two diagrams of Fig.~\ref{fig2}. The extension to the nonlocal case leads
to further diagrams, which need to be included in order to guarantee gauge invariance
(a full discussion can be found in~\cite{Branz:2007xp,Faessler:2003yf}).

The gauge invariant form of the matrix element for the radiative transition in case of real 
photons reads as: 
\eq 
M_{\mu\nu} = ( g_{\mu\nu} q_1 q_2 - q_{2 \mu} q_{1 \nu} ) \, 
g_{H\gamma\gamma} \,, 
\en 
where $q_1$ and $q_2$ are the 4-momenta of the photons. 
For the local case the expression for the effective coupling 
constant $g_{H\gamma\gamma}$ is given by the simple form: 
\eq
g_{H\gamma\gamma} = \frac{g_H}{(4 \pi m \zeta)^2} 
\biggl[ \beta^2(\zeta) - 1 \biggr] \,. \label{eq:Hgg}
\en 
However, this expression becomes more complicated and can only be solved numerically
when including finite size effects (see e.g. Appendix of \cite{Branz:2007xp}).
The radiative decay width of the $0^{++}$ molecule is calculated 
according to the formula  
\eq  
\Gamma(H \to \gamma\gamma) = \frac{\pi}{4} \alpha^2 M^3 
g_{H\gamma\gamma}^2 \,.
\en 
In the local limit we can write the decay width in terms of the quantity $\zeta$ by using
Eq. \eqref{eq:Hgg}
\eq  
\Gamma(H \to \gamma\gamma) = \frac{\alpha^2}{2\pi} \, m  \, I(\zeta) \,, 
\en 
with 
\eq 
I(\zeta) = \zeta \, \sqrt{1-\zeta^2} \, 
\frac{(\beta^2(\zeta) -1)^2}{\beta(\zeta) - \sqrt{1-\zeta^2} } \,. 
\en

Since in the case of hadronic bound states the binding energy $\epsilon$ 
is small in comparison to the masses of the constituent mesons $m$ 
we can perform an expansion of $g_H$, $g_{H\gamma\gamma}$ and 
$\Gamma(H \to \gamma\gamma)$ in $x=\epsilon/(2m)$, 
where $\zeta=1-x$. In order to guarantee an accurate approximation of the 
observables we need to include the leading (LO) and next-to-leading order (NLO) terms 
 in the expansion of $g_H$ and $g_{H\gamma\gamma}$. Therefore, the expansion of $\Gamma(H \to \gamma\gamma)$
includes three terms up to the  next-to-next-to-leading order (NNLO) contribution.

The corresponding $x$-expansions of the quantities of interest 
are given by: 
\eq\label{eq_final} 
& &g_H = m \, I_H(x) \,, \hspace*{.5cm} 
g_{H\gamma\gamma} = m^{-1} \, I_{H\gamma\gamma}(x) \,, \nonumber\\
& &\Gamma(H \to \gamma\gamma) = 
\biggl(\frac{\pi\alpha}{4}\biggr)^2 \, 
m \, J_{H\gamma\gamma}(x)\,,
\en 
where 
\eq 
I_H(x) &=& 8 \, \sqrt{2 \pi} \ (2x)^{1/4} 
\biggl( 1 + \frac{2}{\pi} \sqrt{2x}  
\biggr) + {\cal O}(x^{5/4}) \,, \nonumber\\
I_{H\gamma\gamma}(x) &=& \frac{\sqrt{2 \pi}}{8} \, \ (2x)^{1/4} 
 \biggl( 1 - \frac{4}{\pi^2} - \frac{2}{\pi} 
\biggl(1 + \frac{4}{\pi^2} \biggr) \sqrt{2x} \biggr) \nonumber\\
 &+& {\cal O}(x^{5/4}) \,, \nonumber\\
J_{H\gamma\gamma}(x) &=& \sqrt{2x} \, 
\biggl( \biggl( 1 - \frac{4}{\pi^2} \biggr)^2 - 
\frac{4}{\pi}  \biggl( 1 - \frac{16}{\pi^4} \biggr) \sqrt{2x}  
\nonumber\\ 
&+& 2x \biggl( \frac{7}{8} + \frac{9}{\pi^2} - \frac{50}{\pi^4} 
+ \frac{256}{\pi^6} \biggr) \biggr)  + {\cal O}(x^2) \,. 
\en 
In the next step we consider bound states of the pseudoscalar 
$D$, $D_s$ and $B$ mesons which have the following structure
\eq\label{H_structure} 
|H_D\ra &=& \frac{1}{\sqrt{2}} \big(| D^{+} D^{-} \ra + 
|D^{0} \overline{D^{0}} \ra \big)\,, \nonumber\\ 
|H_{D_s}\ra &=& | D^{+}_s D^{-}_s \ra \,\\ 
|H_B\ra &=& \frac{1}{\sqrt{2}} \big(| B^{+} B^{-} \ra +  
|B^{0} \overline{B^{0}} \ra \big)\,. \nonumber 
\en  
Note, the existence of bound states of two heavy pseudoscalar 
mesons was proposed before in Ref.~\cite{Gamermann:2006nm}
where hadronic molecules are dynamically generated in a
coupled channel formalism. 
Based on the identifications of Eq.~(\ref{H_structure}) additional flavor factors
have to be considered in Eq.~(\ref{eq_final}) which leads to:
\eq\label{eq_final_update} 
& &g_{H_i} = m_i \, I_{H}(x_i) \,, \hspace*{.5cm} 
g_{H_i\gamma\gamma} =    
\frac{c_{H_i\gamma\gamma}}{m_i} \, I_{H\gamma\gamma}(x_i) \,, \nonumber\\ 
& &\Gamma(H_i \to \gamma\gamma) = \biggl(\frac{\pi\alpha}{4}\biggr)^2 \, 
m_i \, c_{H_i\gamma\gamma}^2 \, J_{H\gamma\gamma}(x_i) \,, 
\en 
where $x_i = \epsilon/(2m_i)$, $c_{H_i\gamma\gamma} = 1/\sqrt{2}$ 
for $i = D, B$ and $1$ for $i = D_s$.
From last equation follows the ratio of the two-photon widths of 
two different molecular states which is characterized by 
the respective constituent masses and binding energies. 
At leading order we deal with the simple expression:
\eq 
\frac{\Gamma(H_B \to \gamma\gamma)}{\Gamma(H_D \to \gamma\gamma)} 
\sim \Big(\frac{m_B \epsilon_B}{m_D \epsilon_D}\Big)^{1/2} ~.
\en 
Varying the binding energy $\epsilon$ from $10$ to $100$ MeV which (besides the exception of the X(3872))
are typical values in the heavy meson sector,
the two-photon decay widths are evaluated as
\eq\label{eq_local} 
\Gamma(H_D \to \gamma\gamma) &=& 0.25 - 1.19\ {\rm keV}\,, \nonumber\\ 
\Gamma(H_{D_s} \to \gamma\gamma) &=& 1.31 - 2.58 \ {\rm keV} \,, \\ 
\Gamma(H_B \to \gamma\gamma) &=& 1.19 - 2.78 \ {\rm keV} \,. \nonumber
\en 
Since $\Gamma(H_i\to\gamma\gamma)\propto \sqrt{m\epsilon}$ the smaller value of the binding energy
also results in a smaller radiative decay width.

Up to now the results for the coupling and the two-photon decay widths were given in the local limit.
For the physically appropriate description of an extended object, the hadronic molecule, finite
size effects in terms of the vertex function have to be included. This will lead to a
suppression of the couplings and the two-photon decay widths, hence the local case
presents an upper limit for these quantities.
In the following we study in addition to the influence of mass and binding energy the dependence on the
size parameter $\Lambda_{H_i}$, which models the nonlocality.

\begin{figure}[tb]
\centering{
\hspace*{-.4cm} 
\epsfig{figure=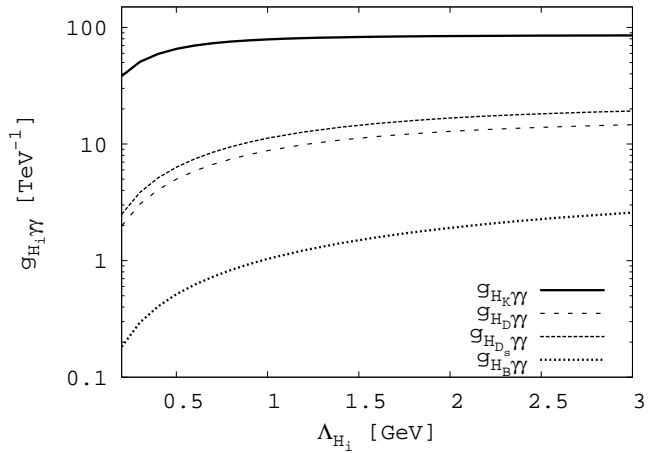,scale=.7}}
\caption{Couplings $g_{H_K\gamma\gamma}$ and $g_{H_i\gamma\gamma}$ with $i=D,D_s,B$ in dependence 
on $\Lambda_{H_i}$ for $\epsilon_K=7.35$ MeV and $\epsilon_{i}=10$ MeV.} 
\label{fig4}
\end{figure}

\begin{figure}[tb]
\centering{
\hspace*{-.4cm} 
\epsfig{figure=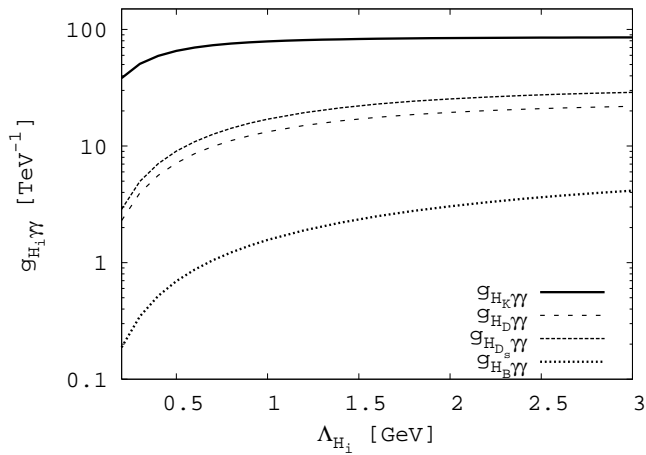,scale=.7}}
\caption{Couplings $g_{H_K\gamma\gamma}$ and $g_{H_i\gamma\gamma}$ with $i=D,D_s,B$ in dependence 
on $\Lambda_{H_i}$ for $\epsilon_K=7.35$ MeV and $\epsilon_{i}=100$ MeV.} 
\label{fig6}
\end{figure}

The dependence of the couplings $g_{H_i\gamma\gamma}$ on the 
size parameter $\Lambda_{H_i}$, typically chosen in the range from 1 to 3 GeV,
is illustrated in the logarithmic plots of Figs.~\ref{fig4}-\ref{fig6}
for different values of $\epsilon$ ranging from 10 to 100 MeV.
For comparison we also include the coupling $g_{H_K\gamma\gamma}$ of $f_0(980)/a_0(980)$ to two
photons with the well determined binding energy of $\epsilon=7.35$ MeV.
The local case, i.e. $\Lambda_{H_i} \to\infty$, characterizes the asymptotics of the curves. 
The convergence of the coupling $g_{H_i\gamma\gamma}$ towards the local or asymptotic
value depends on the constituent meson masses.
The coupling $g_{H_K\gamma\gamma}$ is almost stable with respect to variations of
$\Lambda_{H_i}$ near 1 GeV. The couplings of the heavy hadron molecules
are more sensitive to finite size effects for values near $\Lambda_{H_i}\approx 1$ GeV, note that
in Figs.~\ref{fig4}-\ref{fig6} the dependence on $\Lambda_{H_i}$ is displayed on a log scale.
In fact, the coupling $g_{H_K\gamma\gamma}$ reaches 90\% of the asymptotic value
(local approximation) already at $\Lambda_{H_i}\approx0.6$~GeV. In contrast,
the couplings of heavy bound states approach 90\% of the local value at around 4 GeV in case of
the $D$ and $D_s$-meson bound states and at about 9 GeV for the even heavier hidden-bottom molecule.

The behavior of the couplings $g_{H_i\gamma\gamma}$, which enter quadratically
in the expression for the radiative decay width, shows that the suppression due to the
size parameter $\Lambda_{H_i}$ is larger in case of heavier constituents.
Phenomenologically, the values for $\Lambda_{H_i} $ tend to increase in case of
heavy bound states. This behavior was already observed
in earlier analyses in the framework of meson molecules~\cite{Branz:2009yt,Faessler:2007gv,Branz:2007xp}
but also in case of baryons~\cite{Branz:2010pq}. 
But still, for reasonable values of $\Lambda_{H_i}$ heavier systems 
are strongly influenced by finite size effects as reflected in the 
two-photon coupling $g_{H_i\gamma\gamma}$.

In practice $\Lambda_{H_i}$ should also depend in average on the binding energy.
For instance, a small binding energy would lead to a loose bound state with a more extended structure
than a strongly bound compact state. Therefore $\Lambda_{H_i}$ should decrease with smaller binding energies,
increasing the deviation from the local limit even more.

To quantify the inclusion of finite size effects for the two-photon decay widths
we also give results for two values of $\Lambda_{H_i}$.
For $\Lambda_{H_i} = 1$ GeV and $\epsilon = 10 - 100$ MeV we have
\eq
\Gamma(H_D \to \gamma\gamma) &=&  0.05 - 0.34 \ {\rm keV}\,, \nonumber\\ 
\Gamma(H_{D_s} \to \gamma\gamma) &=& 0.32 - 0.68   \ {\rm keV} \,, \\ 
\Gamma(H_B \to \gamma\gamma) &=&  0.05 - 0.12   \ {\rm keV} \,. \nonumber
\en 
When increasing  $\Lambda_{H_i}$ to 2 GeV the results are by a factor 2$-$3 
larger
\eq 
\Gamma(H_D \to \gamma\gamma) &=& 0.13 - 0.73  \ {\rm keV}\,, \nonumber\\ 
\Gamma(H_{D_s} \to \gamma\gamma) &=& 0.71 - 1.52  \ {\rm keV} \,, \\ 
\Gamma(H_B \to \gamma\gamma) &=& 0.18 - 0.44\ {\rm keV} \,. \nonumber
\en 
These results should be compared to the local limits of Eq.~(\ref{eq_local}).

From previous discussions it should be clear that finite-size effects are quite important
for a quantitative determination of the radiative decay width of heavy hadron molecules.
The same observation holds for molecular states composed of vector
mesons~\cite{Branz:2009yt}, where we showed that for $\Lambda_{H_{D^\ast}}=\Lambda_{H_{D_s^\ast}} = 1 - 2$ GeV 
the radiative widths of the molecules $Y(3940)= \{D^\ast D^{\ast \dagger}\}$ 
and $Y(4140) = \{ D^\ast_s D^{\ast \dagger}_s \}$ are of order of 1 keV for $\Lambda_{H_i}=2$ GeV.

For completeness, we compare the two-photon decays of the heavy systems to the one of the light scalars
$f_0/a_0$.
The stability of the coupling in case of a $K\bar K$ bound state already implies that the radiative
decay width $\Gamma(f(980) \to \gamma\gamma)$ is not sensitive to the cutoff or 
finite-size effects provided $\Lambda_{H_i}$ is above 0.5 GeV.
Here we completely agree with the conclusions of 
Ref.~\cite{Hanhart:2007wa}). 
In particular, in order to reproduce the current data on strong and 
radiative decays of $f(980)$ we fixed the cutoff parameter 
$\Lambda_{H_K}\equiv\Lambda_{f}= 1$ GeV~\cite{Branz:2007xp} with 
\eq
\Gamma(f(980) \to \gamma\gamma) = 0.25 \ {\rm keV} 
\en 
a value which is very close to the result of the local approximation: 
\eq  
\Gamma(f(980) \to \gamma\gamma) = 0.29  \ {\rm keV} \,. 
\en 
Similar results are obtained by the molecular approach in~\cite{Hanhart:2007wa} and \cite{Oller:1997yg}.
However, for smaller values of 
$\Lambda_{f} < 0.6$~GeV the radiative decay width decreases which 
means that such small values of the cutoff parameter $\Lambda_{f}$ 
are unlikely according to present data. 

\begin{acknowledgments}

This work was supported by the DFG under Contract No. FA67/31-2
and No. GRK683. This research is also part of the
European Community-Research Infrastructure Integrating Activity
``Study of Strongly Interacting Matter'' (HadronPhysics2,
Grant Agreement No. 227431), Russian President grant
``Scientific Schools''  No. 3400.2010.2, Russian Science and
Innovations Federal Agency contract No. 02.740.11.0238.

\end{acknowledgments} 

\end{document}